\documentclass[twocolumn]{jpsj3}
\usepackage{txfonts}
\usepackage{bm}
\usepackage{braket}
\title{Material-dependent screening of Coulomb interaction\\
in single-layer cuprates}

\author{Shingo Teranishi\thanks{teranishi@artemis.mp.es.osaka-u.ac.jp}, 
 Kazutaka Nishiguchi, and Koichi Kusakabe}
\inst{
Depatment of Materials Engineering Science, Graduate School of Engineering Science, 
Osaka University, \\
1-3 Machikaneyama-cho, Toyonaka Osaka 560-8531, Japan}

\abst{
To explore material dependence of cuprate superconductors, 
we evaluate effective Coulomb interactions for Hg1201 and Tl1201, 
where Tl1201 having a nearly half value of $T_c$ of 
Hg1201 even at the optimal oxygen concentration. 
Although structures are similar for these superconductors, 
there is an apparent difference in the occupied levels below $E_F$. The characteristic difference in the band structure is correlated 
with oxygen contents in the buffer layer.
By using constrained Random Phase Approximation, 
effective screened Coulomb interactions are estimated for 
HgBa$_2$CuO$_4$ and TlBa$_2$CuO$_5$. The results shows that the value of screened on-site Coulomb interaction in Hg1201 is nearly twice bigger than that in Tl1201. In addition, The eigenvalues of the linearized Eliashberg equation of single-band Hubbard model within FLEX can show apparent difference in $T_c$. When we assume that the twice big screened on-site Coulomb for Hg, the material dependent $T_c$ might be explained. }

\begin{document}
\maketitle

\section{Introduction}
After the first discovery of a cuprate superconductor in 1986,\cite{La-first} 
plenty of cuprate superconductors have been reported.\cite{Y-first,Hg1201-first,Hg1223_1,Hg1223_2} 
Especially Hg-based cuprates are well known because of 
their high transition temperatures. 
For the high-temperature superconductivity (high-$T_c$), 
in Hg-based cuprate crystal, 
the number $n$ of CuO$_2$ planes should be three 
in a basic stacked-layered structure, {\it i.e.} in the unit cell. 
The critical temperature $T_c$ for superconductivity rises up to 135K 
even at ambient pressure 
when a triple-layered Hg-based cuprate ($n=3$) is chosen 
and when it is optimized with respect to its material parameters, 
{\it e.g.} the oxygen content.\cite{Hg1223_1,Hg1223_2}
Although several other compounds have similar superconducting properties, 
$T_c$ becomes maximum by a choice of the material and 
by the adjustment of its internal material parameters. 

To explore a basic mechanism of high-$T_c$, 
we reconsider several common features and detailed difference 
among cuprates, which should be explained by a unique theory. 
One of the special characters of high-$T_c$ 
is material dependence of the superconducting properties. 
There are several known classification of cuprate superconductors in series, 
{\it e.g.} Hg-compounds, Tl-compounds, and Bi-compounds, 
which essentially differ in the atomic structure of the buffer layer. 
When we see some specific materials, and if we compare 
$T_c$ of some compounds in different series, we can actually find  
several hints to understand the relevant superconducting mechanism. 

We see global similarity in doping dependence of the CuO$_2$ planes. 
The optimal doping is often found at around a hole concentration 
of 0.16 per a CuO$_2$ plane. To adjust the concentration, 
one often needs to modify the buffer layers in its oxygen contents. 
When a high-$T_c$ material is optimized with respect to the hole concentration, 
at the optimal doping, the layer-number dependence of 
the transition temperature $T_c$ in several series of cuprates may be derived. 
Comparison among materials categorized in these series was made 
in experiment.\cite{mukuda2011high} 
Triple-layer compounds provide the highest $T_c$ among 
multi-layered compounds in various series. 
In addition, some special features of series dependence were concluded, 
{\it e.g.} existence of a large $n$ regime. 
When the layer number $n$ becomes more than around 7, $T_c$ reaches at 
a saturated value for each series. 
Actually, experimental findings of material dependence in nature 
always extend our understanding of the high-T$_c$ superconductivity. 

Theoretically, there are many successful explanations 
on tendency of the material dependence. 
An example was the Fermi surface shape dependence of 
several cuprate series.\cite{Fermishape_2,Fermishape_1} 
In this direction, Sakakibara, {\it et al.} have explored that 
the single-layer Hg-based cuprate is in a good condition for 
the orbital distillation effect,\cite{Sakakibara2010,Sakakibara2012,Sakakibara2014} 
while the 214 phase of La compounds may be a mixed multi-$d$-band system. 
They proposed that a purified 3d$_{x^2-y^2}$ band for Hg-compounds 
should provides better high-$T_c$, while hybridization of 3d$_{x^2-y^2}$ 
and 3d$_{3z^2-r^2}$ components around the Fermi level may 
causes reduction in $T_c$. Even with this understanding, 
however, there remains unresolved material dependence. 

Here an important hint can be found in the $T_c$ difference 
between the Hg-series and the Tl-series. 
A known experimental fact is that 
$T_c$ of Hg1201 is $T_c\approx 100$K and 
that of Tl1201 is $T_c\approx 50$K at the optimal doping.\cite{mukuda2011high} 
Indeed, $T_c$ of Tl1201 is only a half of the value of Hg1201. 
As we will show, the band structure calculation and the tight-binding fitting 
by the Wannierization technique\cite{wannier1,wannier2} 
tell that relevant 3d$_{x^2-y^2}$ bands 
of these compounds resemble each other. 
%Conversely speaking, a single-band model derived by limited numbers 
%of orbitals may not completely describe the material dependence. 

On this problem, there had been several discussions on the $T_c$ value 
and its dependence on materials parameters. 
The largeness of Madelung potential at the apical oxygen site,\cite{apex}  
the orbital energy difference between 3d$_{x^2-y^2}$ and 
3d$_{3z^2-r^2}$\cite{Sakakibara2010,Sakakibara2012,Sakakibara2014}	, the largeness of interlayer tunneling effect\cite{anderson1995interlayer}, 
the lower amount of disorder on the CuO$_2$ plane were considered. 
As for the former three factors, however, the band structure calculation 
had captured the material characteristics as far as the single-particle 
transfer terms are concerned. 
The third effect should be reconsidered with careful consideration of 
two-particle parts relevant for the interlayer pair-hopping processes 
described by an effective Hamiltonian.\cite{Nishiguchi2013}  
The last point would be experimental. However, we should note 
stiff nature of the cuprate high-temperature superconductivity 
against potential scatters.\cite{Potential_eisaki} 

Here, we propose another factor of the material dependence. We focus on on-site Coulomb interaction and its screening effects. In many cases, Coulomb interaction is treated as material-independent values. However, we propose that on-site Coulomb interaction can strongly depends on oxygen contents in the buffer layer.

In this paper, we analyze the band structures given by 
the generalized-gradient approximation of DFT. 
By comparing the energy bands of Hg-based and Tl-based compounds, and then we apply constrained Random Phase approximation(constrained-RPA) method to evaluate Coulomb interactions. Then, to discuss the different strength of superconductivity in Hg-based and Tl-based compounds, we apply Fluctuation Exchange Approximation(FLEX)  to single-band Hubbard model and solve the linearized Eliashberg equation.
\section{Methods}
\subsection{constrained-RPA}
In this section, we introduce constrained-RPA methods which is recently-developed calculation method\cite{cRPA}. In this method, we divide polarization($P$) into two contributions. One is by transition among target bands($P_d$) and the other is by rest of transitions($P_r$).% The bare Green's function is given by
%\begin{align}
%G_d( r, r';\omega)=\sum^{occ}_d\frac{\psi_d( r)\psi^*_d( r')}{\omega-\varepsilon-i0^+}+\sum^{unocc}_d\frac{\psi_d( r)\psi^*_d( r')}{\omega-\varepsilon+i0^+}.\\
%G_d(\bm r,\bm r';\omega)=\sum^{occ}_d\frac{\psi_d(\bm r)\psi^*_d(\bm r')}{\omega-\varepsilon-i0^+}+\sum^{unocc}_d\frac{\psi_d(\bm r)\psi^*_d(\bm r')}{\omega-\varepsilon+i0^+}.
%\end{align}
%Let $P$ be the total polarization, including the transitions between all bands, 
%\begin{align}
%\nonumber
%P( r, r';\omega)=\sum^{occ}_i\sum^{unocc}_j\psi_i( r)\psi^*_i( r')\psi^*_i( r)\psi_i( r')\\\times\left\{\frac{1}{\omega-\varepsilon_j+\varepsilon_i+i0^+}-\frac{1}{\omega+\varepsilon_j-\varepsilon_i-i0^+}\right\}.
%\end{align}
%where 
%$P$ can be divided into $P=P_d+P_r$, in which $P_d$ includes only 3d to 3d translations, and $P_r$ be the rest of the polarization.
The screened interaction W on the RPA level is given by
\begin{align}
W&=[1-vP]^{-1}v=[1-vP_r-vP_d]^{-1}v\\
&=[(1-vP_r)\{1-(1-vP_r)^{-1}vP_d\}]^{-1}v\\
&=\{1-(1-vP_r)^{-1}vP_d\}^{-1}(1-vP_r)^{-1}v\\
&=[1-W_rP_d]^{-1}W_r.
\end{align}
where we have defined a screened interaction $W_r$ that does not include the polarization from the 3d-3d transitions:
\begin{align}
W_r=[1-vP_r]^{-1}v.
\label{eq:c-rpa}
\end{align}
Screened Coulomb interactions at one time in the Wannier basis is expressed as
\begin{align}
V_{ij}=\int d\bm r\int d\bm r'\phi^*_i(\bm r)\phi_i(\bm r) W_r(\bm r,\bm r')\phi^*_j(\bm r')\phi_j(\bm r').
\end{align}
Here $i$ and $j$ are the indices of the Wannier orbitals. $W_r(\bm r,\bm r')$ is a screened Coulomb interaction.
\begin{align}
W_r(\bm r, \bm r)=\frac{4\pi}{\Omega}\sum_{\bm{{qGG'}}}\frac{e^{-\mathrm{i}(\bm{{q+G}})\bm{{r}}}}{|\bm{{q+G}}|}\epsilon^{-1}_{\bm{{GG'}}}(\bm{{q}})\frac{e^{-\mathrm{i}(\bm{{q+G'}})\bm{\mathrm{r'}}}}{|\bm{{q+G'}}|}
\end{align}
where $\Omega$ is the crystal volume, $\epsilon^{-1}_{\bm{{GG'}}}(\bm{{q}})$ is the inverse dielectric matrix, $\bm q$ is a wave vector in the first Brillouin zone and $\bm G$ is a reciprocal lattice vector. The dielectric matrix is expressed as
\begin{align}
\epsilon_{\bm{{GG'}}}(\bm{{q}})=\delta_{\bm{GG'}}-v(\bm{q+G})\chi_{\bm{GG'}}(\bm q),
\end{align}
Where $v(\bm q)=4\pi/\Omega|\bm q|^2$ is the bare Coulomb interaction. The porlarization matrix in constrained-RPA method is expressed as
\begin{align}
\nonumber
\chi_{\bm{{GG'}}}(\bm q)=\sum_{\bm k}\sum_{\alpha\beta}\bra{\psi_{\alpha\bm{k+q}}}e^{-\mathrm{i}(\bm{q+G})\bm{r}}\ket{\psi_{\beta\bm k}}\\
\times\bra{\psi_{\beta\bm{k}}}e^{-\mathrm{i}(\bm{q+G})\bm{r}}\ket{\psi_{\alpha\bm {k+q}}}\frac{f_{\alpha\bm{k+q}}-f_{\beta\bm k}}{E_{\alpha\bm{k+q}}-E_{\beta\bm k}},
\end{align}
Here $\psi_{\alpha\bm k}$ is the Bloch state, $E_{\alpha\bm k}$ is the energy of the state, and $f_{\alpha\bm k}$ is the occupancy. $\alpha,\beta$ stand for the bands which do not include 3d-3d band transitions.

To perform the calculation and derive effective interactions, we use RESPACK-code\cite{respack1,respack2,respack3,respack4,respack5}%, which is free-source software made by K. Nakamura et al. This code include the wannierization method with the projection operators
%for the disentanglement process\cite{wannier1,wannier2,imada_disentanglement}.
\section{Results}
\subsection{Hg- and Tl-based cuprate compounds}

For comparison, we consider Hg-compounds and Tl-compounds. 
When a cuprate crystal is prepared at an optimal doping, the structure 
often becomes an alloy or a mixed phase. 
To adopt reliable DFT codes\cite{qe,vasp_1,vasp_2,vasp_3} in our simulation, however, 
we need to consider a perfect periodic crystal with a unit cell. 
Therefore, the simulation becomes possible by limiting 
the filling factor at some special value allowing construction of a super cell. 
Owing to this reason, we treat a crystal phase fixing concentration of 
dopant and an oxygen composition ratio. 
In some cases, we look at a filling factor corresponding to 
the half-filling of the CuO$_2$ plane.% Note that the obtained band structure merely 
%gives a spectrum of a single-particle part of the many-body effective Hamiltonian. 
%For a good Mott insulator, therefore, the band structure of this single-particle 
%part should show metallic features rather than an insulating gapped phase. 

%For the filling control, we mainly consider creation of oxygen deficiency. 
%The oxygen concentration is controlled 
%by reduction/oxidation of the buffer layers. 
%Therefore, when we consider, for example, 
%HgBa$_2$CuO$_{4+\delta}$ with $\delta$ from 0 to 1, 
%we treat oxygen deficiency at the HgO$_\delta$ plane maintaining 
%perfectness of CuO$_2$ layers. 
%Similarly, when we consider three-layer $n=3$ compounds, 
%we consider the oxygen deficiency in the buffer layers. 
%For the single-layer structures, we treat 
%HgBa$_2$CuO$_4$ and TlBa$_2$CuO$_5$. 
%The structural parameters for each crystal are determined by 
%the optimization simulation in DFT-GGA, 
%where the external pressure condition is zero. 
%(Fig.~\ref{fig:structure}) 

\begin{figure}[t]
 \begin{minipage}[b]{0.4\linewidth}
  \centering
% \hspace*{+	2em} 	
 \includegraphics[keepaspectratio, scale=0.27]{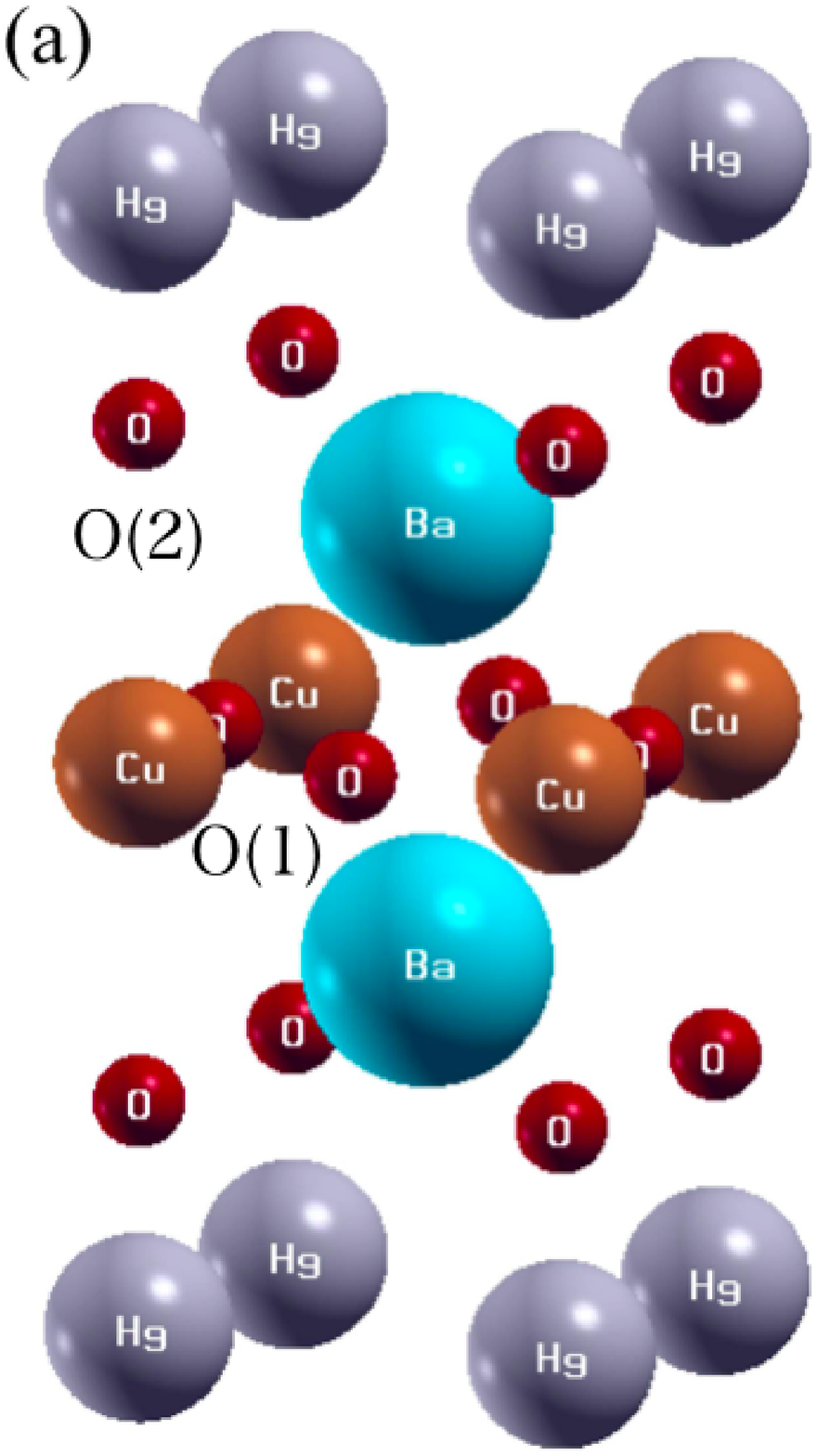}
 \end{minipage}
% \vspace*{2em}
 \begin{minipage}[b]{0.4\linewidth}
  \centering
% \hspace*{2em} 
  \includegraphics[keepaspectratio, scale=0.26]{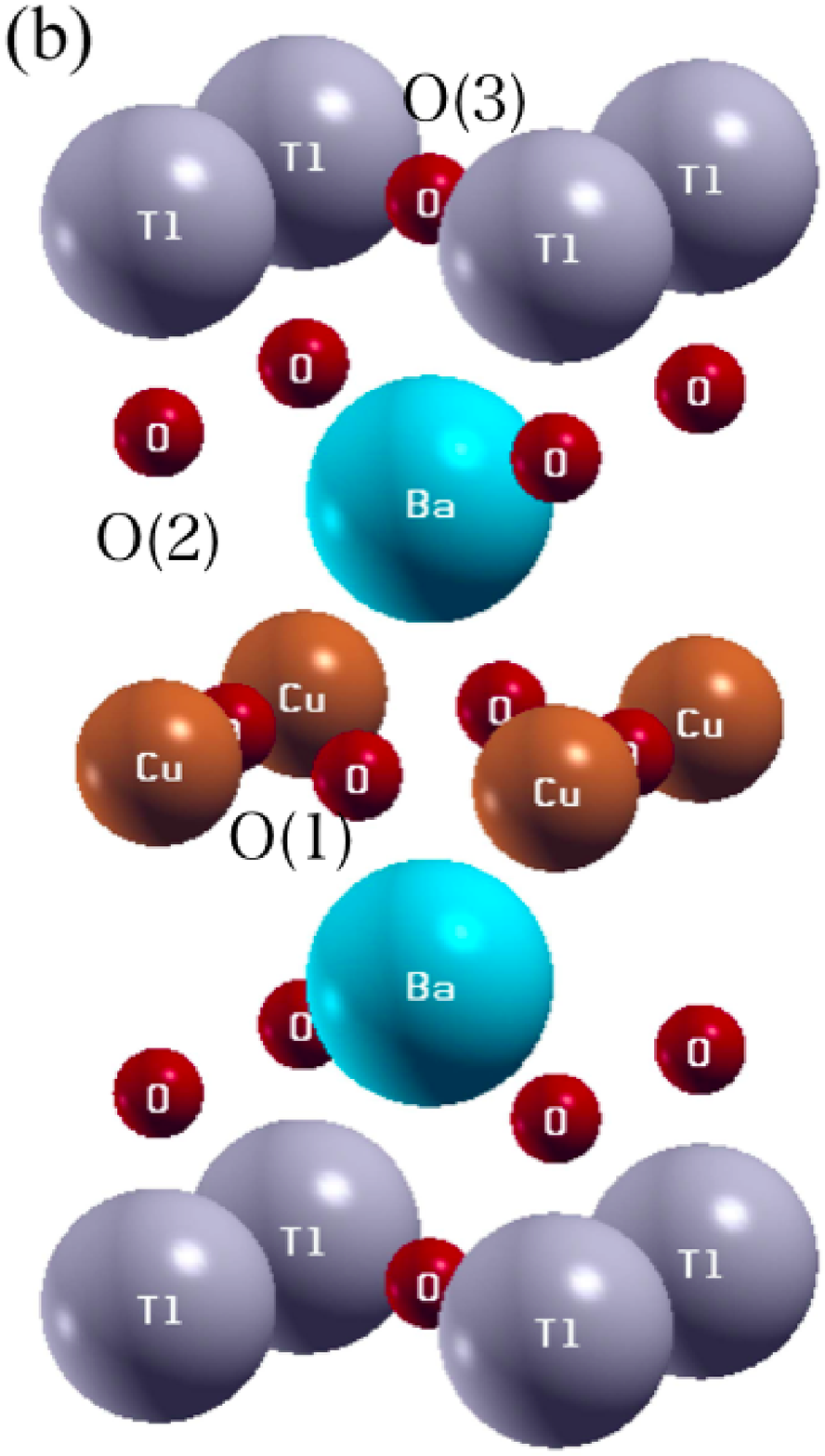}
 \end{minipage}
\caption{The atomic structures of (a) HgBa$_2$CuO$_4$(left) and  
(b)TlBa$_2$CuO$_5$(right). \label{fig:structure}There are three different oxygen in a unit cell. In this figure, O(1) is oxygen in copper-oxygen layer, O(2) is apical oxygen and O(3) is oxygen in the buffer layer.}
\end{figure}

HgBa$_2$CuO$_4$(Hg1201) lacks oxygen atoms at each HgO$_\delta$ plane. 
(Fig.~\ref{fig:structure} (a)) 
We have a local OHgO structure along the $c$ axis.  
Oxygen atoms in this OHgO structure may be 
interpreted as apical oxygen atoms of CuO$_4$ pyramids. 
While, TlBa$_2$CuO$_5$(Tl1201) has TlO planes. (Fig.~\ref{fig:structure} (b)) 

%%%%%%%%%%%%%
%%%%%%%%%%%%%%
%%%%%%%%%%%%%%%%%%%%%%%%%%
 %\vspace*{-4em} 
\begin{figure}[t]
 \begin{minipage}[b]{0.32\linewidth}
  \hspace*{-5em} 	
  \includegraphics[keepaspectratio, scale=0.8]{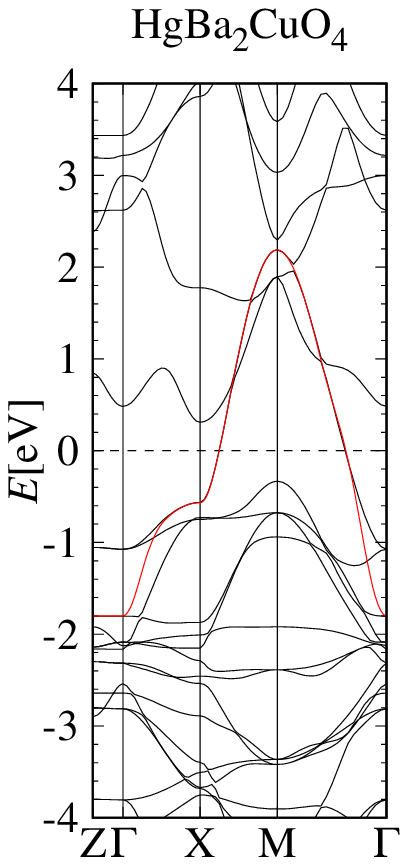}
 \end{minipage}
 \begin{minipage}[b]{0.32\linewidth}
  \hspace*{-2em} 
  \includegraphics[keepaspectratio, scale=0.8]{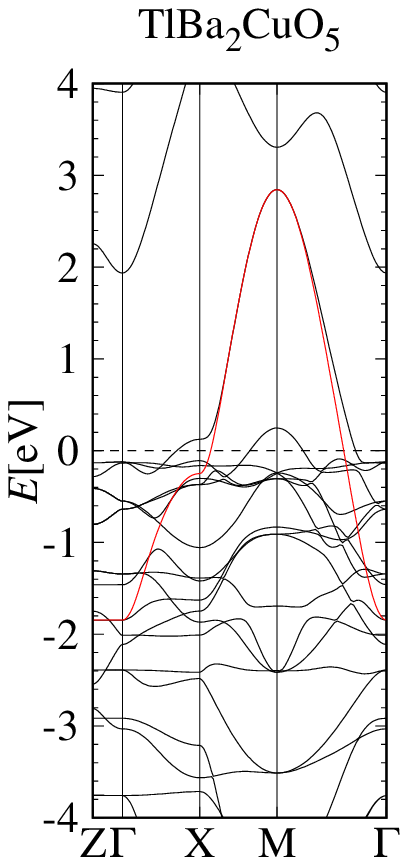}
 \end{minipage}
% \vspace*{+4em} 
\caption{Band structures of HgBa$_2$CuO$_4$ (left) and TlBa$_2$CuO$_5$ (right). Wannier interpolated bands are plotted with red line. Here the Fermi level is set to 0.\label{fig:band_single}}
\end{figure}
%%%%%%%%%%%%%%%%%%%%%%%%%
A nominal filling factor in the CuO$_2$ plane 
may be counted by the rule for ionization valence of 
noble metal, alkaline earth metal ions, and oxygen. 
Supposing Hg$^{+2}$, Tl$^{+3}$, Ba$^{+2}$, and O$^{-2}$, 
the Cu formal valence for these compounds are 
+2 (half-filling) for HgBa$_2$CuO$_4$ and 
+3 (one hole doped per unit cell) for TlBa$_2$CuO$_5$. 

%\subsection{Single-layer cuprate superconductors}
%\label{single-layer}
We show the band structures of HgBa$_2$CuO$_4$, and  
TlBa$_2$CuO$_5$ in Fig.~\ref{fig:band_single}.
In this calculation, we utilize the norm-conserving pseudo potentials 
with the PBE functional. The energy cut-offs in the plane-wave expansion for 
the wave function and the charge density are (100, 400) [Ry]. 
For the self-consistent charge density construction, 
the integration with respect to the $k$ vectors is done 
using a 8$\times$8$\times$8 $k$-point mesh in the 1st Brillouin zone. 
The unit cell of these compounds are optimized in the simulation, 
where the pressure control is done with 
a criterion of each diagonal element of the stress tensor being less than 0.5 [kbar]. 
The internal atomic structures 
are optimized with a criterion that the summation of the absolute values 
of force vector elements becomes smaller than $1.0\times 10^{-8}$[Ry/a.u.]. 

We show the hopping parameters for the 3d$_{x^2-y^2}$ band in Hg1201,Tl1201 in Table. \ref{table:hopping}. Here, a little larger value of $t$ for the Tl-compound comes 
from its shrunk lattice constant. When we use the lattice constant 
of Tl1201 for Hg1201, the value of $t$ becomes close to -0.57. We can see that the general form of the band structures are almost same. However, Tl1201 has more dense bands below the Fermi energy than Hg1201 has.
%Considering change in $T_c$ in high pressure for Hg-compound, 
%the difference in the transition temperature can not be fully explained by 
%the in-plane transfer parameters. 
%Except when the effective screened interaction 
%$U$ is strongly material dependent, it is not easy to derive 
%the reduction of $T_c$ for Tl1201 
%by modeling with a single-band Hubbard Hamiltonian. 
%We find that the oxygen bands coming from CuO$_2$ plane 
%are also similar for these compounds. 
%This is rather natural, because difference in the material structures 
%among Hg1201 and Tl1201 compounds 
%comes essentially from the buffer layers. The difference in the in-plane 
%lattice constant is not enough large to modify the quantitative values 
%of intra-layer hopping parameters. 
%%%%%%%%%%%%%%%%%%%%%%%%%
\begin{figure}[t]
% \vspace*{-10em} 
 \begin{minipage}[b]{\linewidth}
% \hspace*{+5em} 	
  \includegraphics[keepaspectratio, scale=0.5]{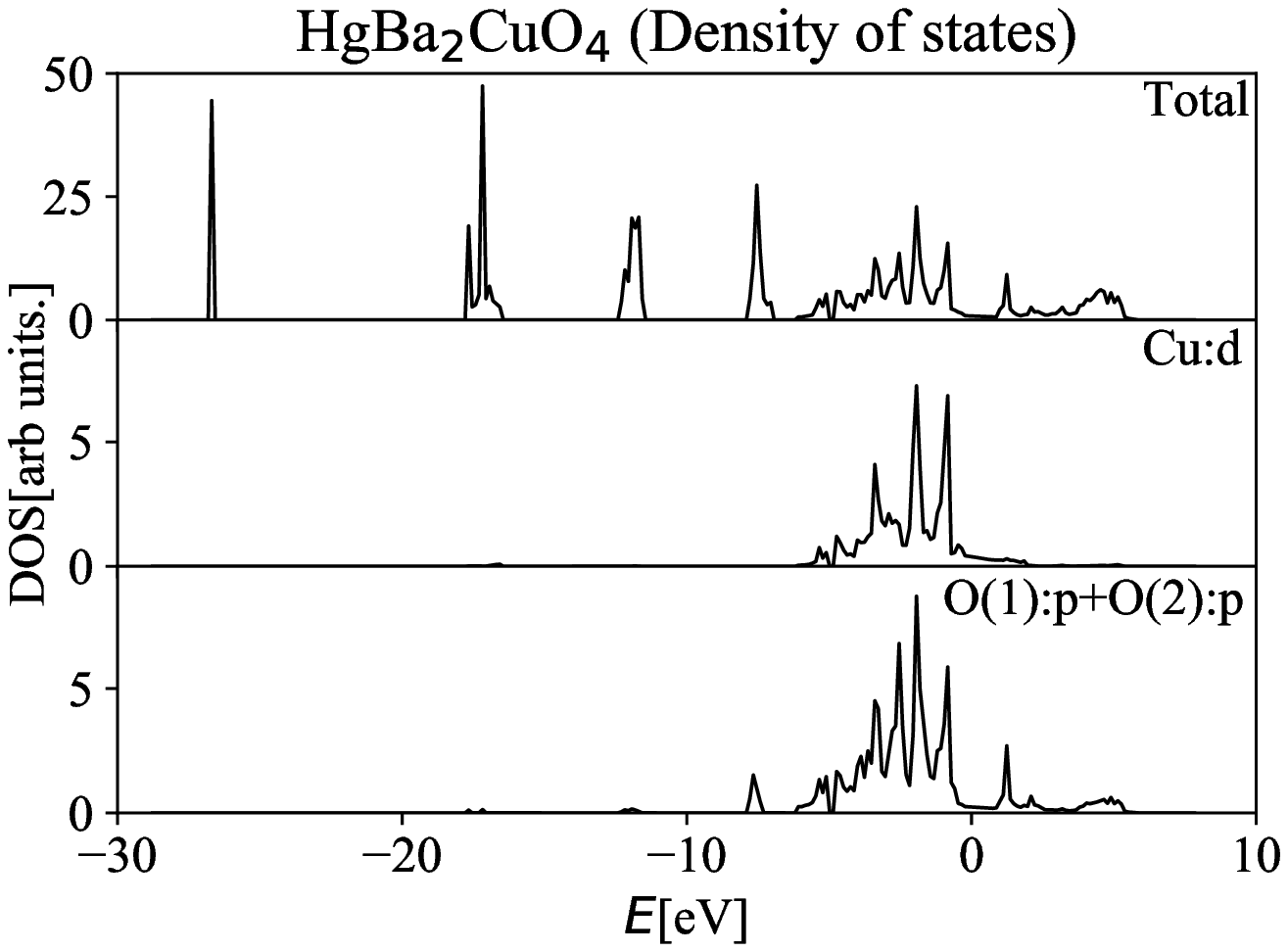}
 \end{minipage}\\
 \begin{minipage}[b]{\linewidth}
% \hspace*{+5em} 
  \includegraphics[keepaspectratio, scale=0.5]{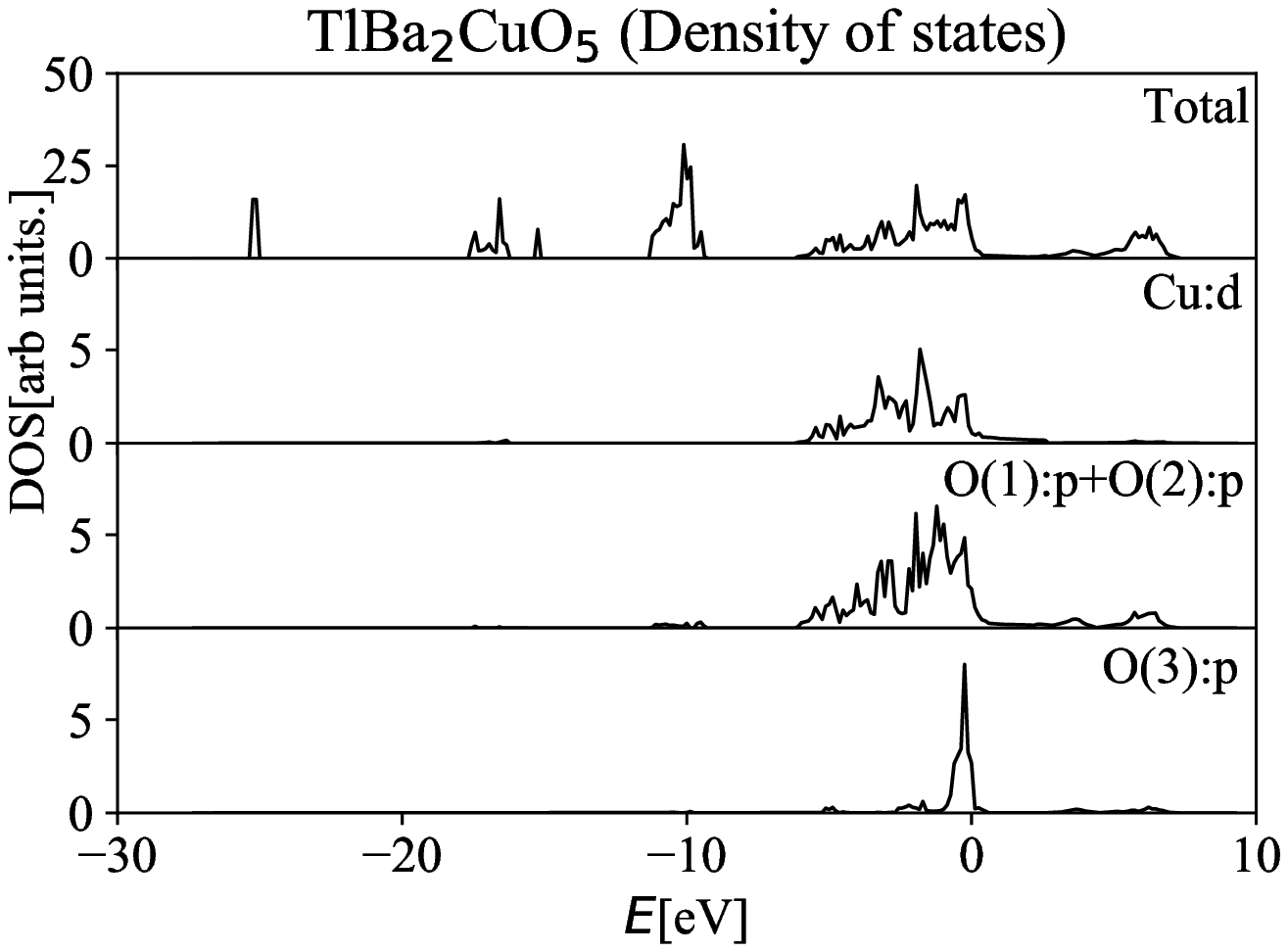}
 \end{minipage}
%\vspace*{+12em} 
\caption{Density of states of HgBa$_2$CuO$_4$ (upper) and TlBa$_2$CuO$_5$ (lower). Here the Fermi level is set to 0.\label{fig:dos_single} In Hg1201, we show partial density of states onto d orbitals of copper (Cu:d), p orbitals of oxygen in copper-oxygen plane and apical site (O(1):p+O(2):p). In addition to it, we show partial density of states onto p orbitals of oxygen in the buffer layer (O(3):p) in Tl1201.}
\end{figure}
%%%%%%%%%%%%%%%%%%%%%%%%%

 Here we also show the total density of states and partial density of states onto orbitals and atoms. According to the results in Fig. \ref{fig:dos_single}, p orbitals of oxygen in the buffer layer(O(3):p) give large contribution to density of states in the occupied levels slightly below $E_F$ in Tl1201.% From analysis of crystal structure, we found that the contribution comes from oxygen in buffer layer. 
%%%%%%%%%%%%%%%%%%%%%%%%%%
\begin{table}[h]
  \caption{The hopping parameters for the 3d$_{x^2-y^2}$ band in Hg1201,Tl1201.
  \label{table:hopping}}
  \centering
  \begin{tabular}{lccc}
    \hline
& Hg1201  &  Tl1201  \\
    \hline \hline
  $t [\mathrm{eV}]$  &  -0.450 &  -0.574 \\
    $t' [\mathrm{eV}]$  &   0.102 & 0.0919 \\
   $t'' [\mathrm{eV}]$  &  -0.095 & -0.0764 \\
     \hline 
  \end{tabular}
\end{table} 
%%%%%%%%%%%%%%%%%%%%%%%%%%
\subsection{constrained-RPA and model calculation}
We show the calculation results of on-site Coulomb interaction int Table. \ref{table:U}. In the course of calculation, the wave function and the charge density are (100, 400) [Ry] and cutoff energy for polarization functions are 10 [Ry]. We use 8$\times$8$\times$8 $k$-point mesh in the 1st Brillouin zone and take 100 bands into account in all calculations.
According to Table. \ref{table:U}, the amplitudes of screened on-site Coulomb interaction($U_{\mathrm{screened}}$) are drastically smaller than bare on-site Coulomb interaction($U_{\mathrm{bare}}$) in both cuprates. The values of $U_{\mathrm{bare}}$ are not so different between the two cuprates. On the other hand, the value of $U_{\mathrm{screened}}$ in Hg1201 is nearly twice bigger than that in Tl1201. We also calculate on-site Coulomb repulusion in CaCuO$_2$ which is mother compounds of infinite-layer cuprates and do not have the buffer layer\cite{Azuma1992}. We found that $U_{\mathrm{screened}}$ in  CaCuO$_2$ is also much larger than that in Tl1201, which indicate that the buffer layer has an important role of deciding the reduction of $U_{\mathrm{screened}}$.
%%%%%%%%%%%%%%%%%%%%%%%%%%
\begin{table}[h]
  \caption{The parameters of effective Coulomb interactions for the 3d$_{x^2-y^2}$ band in Hg1201,Tl1201 and CaCuO$_2$.
  \label{table:U}}
  \centering
  \begin{tabular}{lccc}
    \hline
& Hg1201  &  Tl1201 & CaCuO$_2$ \\
    \hline \hline
     $U_{\mathrm{bare}} [\mathrm{eV}]$  &  12.2783 & 13.7685 & 13.7885 \\
     $U_{\mathrm{screened}} [\mathrm{eV}]$  & 2.9462 & 1.7088 & 3.0397 \\ 
%    $U_{\mathrm{bare}} [\mathrm{eV}]$ & 13.6658 & 13.7685 \\
%   $U_{\mathrm{screened}} [\mathrm{eV}]$  & 3.3074 & 1.7088 \\ 
     \hline 
  \end{tabular}
\end{table} 
%%%%%%%%%%%%%%%%%%%%%%%%%%
%\subsection{Model calculation}

To discuss the strength of superconductivity, we introduce the effective single-band Hubbard model. The single-band Hubbard Hamiltonians is
\begin{align}
H=\sum_{ij\sigma}[t_{ij}c^\dag_{i\sigma}c_{j\sigma}+H.c.]+U\sum_i n_{i\uparrow}n_{i\downarrow},
\end{align}
Here $c^\dag_{i\sigma}$ ($c_{i\sigma}$) stands for creation (annihilation) operator for electrons of site $i$ and spin $\sigma$. $n$ represent particle number operators for site $i$ and spin $\sigma$.
To discuss the strength of superconductivity, we perform  the so-called Fluctuation Exchange Approximation (FLEX)\cite{FLEX_1,FLEX_2} with different amplitude of $U$. We solve the linearized Eliashberg equation
\begin{align}
\lambda\Delta(k)=-\frac{k_BT}{N}\sum_{k'} V(k-k')G(k')G(-k')\Delta(k'),
\end{align}
Here $\Delta(k)$ is a gap function, $G(k)$ the dressed Green's function. $\lambda$ is the eigenvalue of the Eliashberg equation, where $\lambda=1$ corresponds to $T=T_c$, so $\lambda$ serves as a measure of the strength of superconductivity. $V(k)$ the effective interaction for spin single paring is represented as 
\begin{align}
V(q)=\frac{3}{2}U^2\chi^{\mathrm{sp}}(q)-\frac{1}{2}U^2\chi^{\mathrm{ch}}(q)
\end{align}
Where $\chi^{\mathrm{sp}}, \chi^{\mathrm{ch}}$ stand for spin susceptibility and charge susceptibility, respectively.
In this calculation, we take $N = 128^2$ sites with 2048 Matsubara’s frequencies, $k_BT = 0.03$ [eV], filling factor $n$ is $n=0.425$ (When $n=0.5$, the system is half-filling). We take 
approximated $U$ in Hg1201 as $|U/t| = 2.9462/0.450 \simeq 6.5$ and $U$ in Tl1201 as $|U/t| = 1.7088/0.574 \simeq 3.0$. We use the hopping parameters for Hg- and Tl-compounds  derived from the band structures based on the density functional theory in Table. \ref{table:hopping}.
%In both cuprates, we use same hopping paramters: $(|t/t'|,|t/t''|)=(0.2,0.16)$ eV.

\begin{figure}[t]
%\vspace*{-15em} 
%\hspace*{+5em} 
 \begin{minipage}[b]{\linewidth}
  \centering
  \includegraphics[keepaspectratio, scale=0.6]{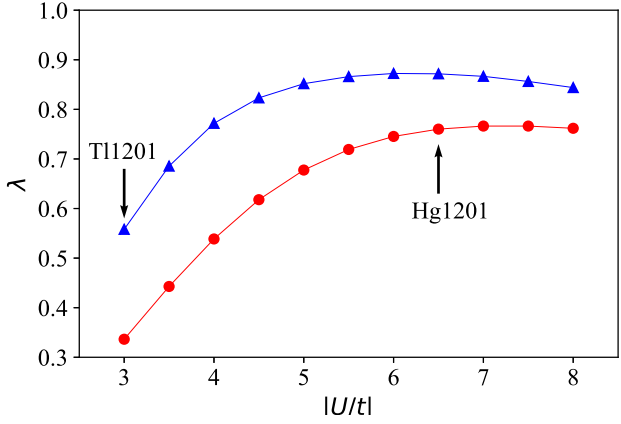}
 \end{minipage}
%\vspace*{+15em} 
\caption{The eigenvalues $\lambda$ of the linearized Eliashberg equation with different amplitude of $U$ in single-band Hubbard model. Red(blue) lines represents the case when hopping parameters for Hg1201(Tl1201) derived from the band structures based on the density functional theory are used. In addition, the arrows indicate the value of $\lambda$s when we use the evaluated $U_{\mathrm{screened}}$ in constrained-RPA.\label{fig:FLEX}}
\end{figure}

 In Fig. \ref{fig:FLEX}, we show the calculated eigenvalues $\lambda$ of the linearized Eliashberg equation the by using FLEX calculation. This results suggests that $\lambda$ becomes larger in Tl1201 than $\lambda$ in Hg1201 when we assume same $U$ for both cuprates, but when we use the evaluated $U_{\mathrm{screened}}$ in constrained-RPA, the expected value of $\lambda$s tell us that the strength of superconductivity in Hg1201 become larger than that in Tl1201. In this calculation, we neglect off-site Coulomb interaction or exchange interactions. However, taking these material-dependent on-site Coulomb interaction into account must be very important to explain the $T_c$ value and its dependence on materials parameters.
%%%%%%%%%%%%%
%%%%%%%%%%%%%%

\section{Discussion} 
\label{discussion_section}
%In addition, we consider lattice constants is changed by the metal-oxygen in buffer layers. 
In Tl-compounds, it is not so easy to access 
the optimal doping by the reduction of oxygen in the buffer layer. 
In a real Tl superconductor, La substitution was used to adjust the filling of CuO$_2$. 
In the real material of TlBa$_{1-x}$La$_x$CuO$_5$, it is known that $T_c$ does not reach the value over 50K even by adjusting the hole concentration.\cite{Tl1201,subramanian1990tlba1} Thus, we may propose to consider a careful control of oxygen in Tl1201 without La doping. If we can adjust the Fermi energy by the reduction of oxygen in the buffer layer, screening effect might be reduced and it will lead to make Coulomb interactions stronger.  Therefore, this procedure might make an enhancement of $T_c$. Actually, this way of approach is consistent with some reported facts.\cite{subramanian1994stabilization} 

In this paper, we mention that metal oxygen in the buffer layers plays an important role to determine the strength of electron-electron interactions, consequently strength of superconductivity. we formed a tendency of material dependence in the screening. When we need to have a strong screened $U$, it is better to make a choice of a divalent metal atom as an element of the buffer layer. This consideration request us to keep essential material structure, i.e. the crystal symmetry. In this respect Cd or Cn in the place of Hg are candidates. Actually there are several reports on similarity in Cd compounds\cite{cd_1,cd_2}. we might be able to say that if additional characters not appear, other dievalent metal atoms in oxides without spin moments, e.g. Hf and W may work. As for as we know, there was no good report on these atoms for replacement. This may come from a reason that there are many atomic sites in cuprates. For example Ca and Ba are often needed. It may not so easy to maintain a crystal symmetry when replacement is done preferably only at Hg site in real materials.

To estimate the screened interaction, we utilized the constrained-RPA method. This method is formulated within RPA for the screened interaction $W$, where the function form of $W_r$ is justified in this approximation. With respect to this limitation, there might be a criticism on the use of $W_r$ for the effective interaction $U$ of the correlated electron model, {\it i.e.} the Hubbard Hamiltonian. Here, we open another route to approach this formulation. Introduction of correlation effects in a generalization of
the  density functional theory (DFT) as the multi-reference DFT (MR-DFT\cite{mr-dft})
is possible by usage of a projection operator $P_A$. This operation
projects a full correlated state $|\Psi\rangle$ in multi-Slater determinants onto
a reference state $P_A|\Psi\rangle$ with quantum entanglement only in
a correlated $d$ band formally without any approximation.
With this step, we can derive a series of correlated electron models,
{\it e.g.} a multi-band extended Hubbard model. When we introduce
a projection $P_A$ on $d$ bands, and $P_B=\hat{1}-P_A$, for example,
we have the direct interaction in the $d$ band $P_A v P_A=v_{dd}$
and the higher order contributions. As a result, we have the next expression
for the full effective screened interaction.
%An approximated expression for an effective screening is obtained via analysis of a rigorous expression for an MR-DFT up-conversion model with the super process. Let $P_A$ be the projection operator on $d$ bands and $P_B = \hat{1} -P_A$.For example, $P_AVP_A = V_{dd}$ is the direct interaction in the d band. We have,
\begin{align}
\nonumber
v_{\mathrm{eff};dd}(\omega)&=P_AvP_A+P_AvP_BG^{(2)}(\omega)P_BvP_A\\
&\simeq v_{dd}+v(P_{(c,d)}(\omega)+P_{(d,v)}(\omega))W_r(\omega)=W_r(\omega).
\end{align}
The approximation leads us to reach an expression similar to constrained-RPA. Let’s suppose that resulted $W_r(\omega)$ is used for a d-band model, so that we can omit subscripts in $v_{dd}$.
\begin{align}
W_r(\omega)=[1-v(P_{(c,d)}(\omega)+P_{(d,v)}(\omega))]^{-1}v.
\end{align}
In this expression, a partial polarization function is given by an effective screened on-site interaction. The term comes from a quantum fluctuation contribution prohibiting fluctuation in the occupation number of the $d$ band.
\begin{align}
\nonumber
P_{(c,d)}(\bm r,\bm r';\omega)=2\sum_{i\in d}\sum_{j\in c_{\mathrm{unoccupied}}}\phi^*_i(\bm r)\phi_j(\bm r)\phi^*_j(\bm r')\phi_i(\bm r')\\
\times\left[\frac{1}{\omega-E_j-U_{\mathrm{eff}}+E_i+i\delta}-\frac{1}{\omega+E_j+U_{\mathrm{eff}}-E_i-i\delta}\right]
\label{eq:mr-dft_1}
\end{align}
\begin{align}
\nonumber
P_{(d,v)}(\bm r,\bm r';\omega)=2\sum_{i\in v_{\mathrm{full-occupied}}}\sum_{j\in d}\phi^*_i(\bm r)\phi_j(\bm r)\phi^*_j(\bm r')\phi_i(\bm r')\\
\times\left[\frac{1}{\omega-E_j-U_{\mathrm{eff}}+E_i+i\delta}-\frac{1}{\omega+E_j+U_{\mathrm{eff}}-E_i-i\delta}\right]
\label{eq:mr-dft_2}
\end{align}
Here, notations $c$ and $v$ represents unoccupied conduction bands and fully filled valence bands. Therefore, in Eq. (\ref{eq:mr-dft_1}), $E_j +U_{\mathrm{eff}} > E_i$, where $U_{\mathrm{eff}}$-contribution is owing to reduction in the number of electrons in d bands. Namely, fluctuation is prohibited by $U_{\mathrm{eff}}$ so that $E_j-E_i+U_{\mathrm{eff}}$. While in Eq. (\ref{eq:mr-dft_2}), we have $E_i +U_{\mathrm{eff}} > E_j$. In this process, the d band is partially doubly occupied. The energy denominator is determined by the condition $E_i-E_j+U_{\mathrm{eff}}>0$.
Appearance of $U$ in the denominator of each expression of $P_{(c,d)}$ and $P_{(d,v)}$ represents
the final-state correlation effect. In the expression of the super process, we have the
two-particle Green function for the expression of the polarization function. Since we
have change in the d occupation in each high-energy process, the energy denominator
is affected by the quantum fluctuation. In a correlated d band, the consistency in the
above expression is certified by
\begin{align}
U_{\mathrm{eff}}=\int d\bm r\int d\bm r'\phi^*_i(\bm r)\phi_i(\bm r)W_r(\bm r,\bm r';\omega=0^+)\phi^*_i(\bm r')\phi_i(\bm r').
\end{align}
This is an ansatz in our proposed procedure. When we replace $P_{(c,d)} + P_{(d,v)}$ with $P_r$ as an approximation,
if we apply $U_{\mathrm{eff}}\sim 0$ in the denominators of Eqs. (\ref{eq:mr-dft_1}) and (\ref{eq:mr-dft_2}),
we arrive at the expression Eq. (\ref{eq:c-rpa}) by constrained-RPA.
\section{Summary and conclusions}
By applying the constrained-RPA method, we evaluated an effective Coulomb interaction for Hg and Tl compounds. As a result, we have an indication of stronger screening in the on-site correlation in a Tl compound relative to Hg compound. Analyzing the nature of the electronic band structures of 
these compounds, we found that there is apparent difference in the occupied levels below $E_F$. These differences reflect the varying crystal structures. Especially, we paid attention to different electronic structures originated from oxygen contents in buffer layer. From analyzing partial density of states and evaluating on-site Coulomb interaction in CaCuO$_2$, we deduce that oxygen contents in the buffer layer have an large effects on screening of Coulomb interactions. However, quantitative evaluation of orbital-decomposed screening effects is future problem.% In addition, there are several factors which determine the strength of screening.% Therefore, calculation is not straightforward. 

 The eigenvalue of the linearized Eliashberg equation within FLEX showed that the estimated difference in $U_{\mathrm{screened}}$ for Hg and Tl can show apparent difference in $T_c$. When we assume that the twice big $U_{\mathrm{screened}}$ for Hg, the material dependent $T_c$ might be explained. %However, off-site Coulomb repulsion may require reconsideration.
 
We also comment on an another effect from the buffer layers. we paid attention to different electronic structure above the Fermi energy. By using MR-DFT method, we evaluated an effective exchange scattering amplitude for some Hg and Tl compounds. Analyzing the nature of the electronic band structures of these compounds, we found that the high energy levels originated from the Hg-O buffer layer contributes well creating enhancement of exchange interaction, which should lead the increase in $T_c$ via the spin fluctuation mechanism. In Tl compounds, however, the enhancement is not so apparent. This is because of the absence of the Tl-O branch around $E_F$, which also 	may explain strange difference between $T_c$s of Hg- and Tl-compounds\cite{2017arXiv170406449T}.
%\begin{itemize}
%\item Tlのshift 
%\item Hgのwannierの取り方
%\item MR-DFT
% \item DOS-VASP-wien2k
% \item c-RPA　m-RPA c-GW
% \item long-ranged
%\end{itemize}
\begin{acknowledgment}
%\acknowledgment
The calculations were done in the computer centers of Kyushu University 
and ISSP, University of Tokyo. 
The work is supported by joint-project for 
``Study of a simulation program for the 
correlated electron systems'' with Advance-soft co. J161101009, 
and JSPS KAKENHI Grant Numbers JP26400357. 
\end{acknowledgment}

%\begin{theliography}{9}
%\bibitem{jpsj} The abbreviation for JPSJ must be ``J. Phys. Soc. Jpn." \note{in the reference list}.
%\end{thebibliography}

\bibliographystyle{jpsj}
\bibliography{c-RPA_cuprates}

\end{document}